\documentclass[conference, 10pt]{IEEEtran}
\IEEEoverridecommandlockouts
\usepackage[table]{xcolor}
\usepackage{latexsym}
\usepackage{graphicx}
\usepackage{amsfonts,amssymb,amsmath}
\usepackage{textcomp}
\usepackage{gensymb}
\usepackage{siunitx}
\usepackage{tikz}
\usepackage[T1]{fontenc}
\usepackage[utf8]{inputenc}
\usepackage{pgfplots}
\usepackage{grffile}
\usepackage{pgfgantt}
\usepackage{pdflscape}
\pgfplotsset{compat=newest} 
\pgfplotsset{plot coordinates/math parser=false}
\usetikzlibrary{plotmarks}
\usetikzlibrary{arrows.meta}
\usepgfplotslibrary{patchplots}
\usetikzlibrary{chains, arrows, calc, positioning}
\usepackage{etoolbox}
\usepackage{multirow}
\usepackage{nicefrac}
\usepackage{optidef}
\usepackage{hhline}
\usepackage{interval}
\usepackage[font=scriptsize]{subcaption}
\usepackage[font=footnotesize]{caption}
\usepackage{xurl}

\def\nb0{{\mathbf{0}}}
\def\nb1{{\mathbf{1}}}

\usepackage[acronyms,nonumberlist,nopostdot,nomain,nogroupskip]{glossaries}
\newacronym{5g}{5G}{5th generation}
\newacronym{6g}{6G}{6th generation}
\newacronym{ncr}{NCR}{network-controlled repeater}
\newacronym{gnb}{gNB}{next-generation nodeB}
\newacronym{ue}{UE}{user equipment}
\newacronym{pa}{PA}{power amplifier}
\newacronym{bh}{BH}{backhaul}
\newacronym{ac}{AC}{access}
\newacronym{rep}{Rep}{repeater}
\newacronym{sc}{SC}{small cell}
\newacronym{mc}{MC}{macro cell}
\newacronym{ris}{RIS}{reflective intelligent surface}
\newacronym{nes}{NES}{network energy saving}
\newacronym{3gpp}{3GPP}{3rd generation partnership project}
\newacronym{mmw}{mmWave}{millimeter-wave}
\newacronym{ntx}{NTx}{transmit antenna}
\newacronym{nrx}{NRx}{receive antenna}
\newacronym{bw}{BW}{bandwidth}
\newacronym{itu}{ITU}{international telecommunication union}
\newacronym{iab}{IAB}{integrated access and backhaul}
\newacronym{capex}{CAPEX}{capital expenditure}
\newacronym{opex}{OPEX}{operational expenditure}
\newacronym{af}{AF}{amplify-forward}
\newacronym{tdd}{TDD}{time-division duplexing}
\newacronym{bf}{BF}{beamforming}
\newacronym{se}{SE}{spectral efficiency}
\newacronym{ee}{EE}{energy efficiency}
\newacronym{qos}{QoS}{quality of service}
\newacronym{snr}{SNR}{signal-to-noise ratio}
\newacronym{eirp}{EIRP}{effective isotropic radiated power}
\newacronym{dl}{DL}{Downlink}
\newacronym{trp}{TRP}{total radiated power}
\newacronym{ru}{RU}{radio unit}
\newacronym{lna}{LNA}{low noise amplifier}
\newacronym{ai}{AI}{artificial intelligence}
\newacronym{iot}{IoT}{internet of things}
\newacronym{fr}{FR}{frequency range}

\newtoggle{conf}
\toggletrue{conf}

\def\BibTeX{{\rm B\kern-.05em{\sc i\kern-.025em b}\kern-.08em T\kern-.1667em\lower.7ex\hbox{E}\kern-.125emX}}

\newcommand{\FigSize}{0.35}

\begin{document}

\title{Towards Energy- and Cost-Efficient \\ 6G Networks}

\author{
\IEEEauthorblockN{Tommy Azzino, Aria HasanzadeZonuzy, Jianghong Luo, Navid Abedini, Tao Luo}
\IEEEauthorblockA{Qualcomm Technologies, Inc. - e-mail: \{tazzino, ahasanza, jianghon, navida, tluo\}@qti.qualcomm.com}
}
\vspace{-5mm}


\maketitle
\begin{abstract}

As the world enters the journey toward the \gls{6g} of wireless technology, the promises of ultra-high data rates, unprecedented low latency, and a massive surge in connected devices require crucial exploration of \gls{nes} solutions to minimize the carbon footprint and overall energy usage of future cellular networks. On the other hand, \glspl{ncr} have been introduced by \gls{3gpp} as a cost-effective solution to improve network coverage. However, their impact on network power consumption and energy efficiency has not been thoroughly investigated. This paper studies \gls{nes} schemes for next-generation \gls{6g} networks aided by \glspl{ncr} 
and proposes optimal \gls{nes} strategies aiming at maximizing the overall energy efficiency of the network.
Repeaters are shown to allow for power savings at \gls{gnb}, and offer higher overall \gls{ee} and \gls{se}, thus providing an energy-efficient and cost-efficient alternative to increase the performance of future \gls{6g} networks.

\end{abstract}
\glsresetall


\IEEEpeerreviewmaketitle

\vspace{-4mm}
\section{Introduction}

With the \gls{6g} of cellular networks on the horizon, we are on the precipice of yet another groundbreaking era. The relentless growth in mobile data traffic, driven by emerging applications and the proliferation of connected devices, sets the stage for the necessity of \gls{6g}. According to forecasts from \gls{itu}, global mobile data traffic is projected to reach impressive levels by 2030, necessitating novel approaches to wireless communication \cite{itu}.
Besides the elevated ambitions of \gls{6g} \cite{6g_vision}, a fundamental question remains: Can we accomplish this technological marvel while being environmentally responsible?

Future \gls{6g} communication systems are foreseen to massively exploit the \gls{mmw} frequency band to enable capacity-eager applications, e.g. autonomous driving and extended reality. 
However, propagation at \gls{mmw} frequencies suffers from severe path loss and blockage, limiting the coverage of the network in these bands, especially in dense urban environments. To overcome these challenges, densification of \gls{sc} systems with massive antenna arrays becomes necessary, resulting in higher \gls{capex} for cellular operators and service providers \cite{road_6g}.

Among other wireless backhauling/fronthauling technologies, such as \gls{ris} and \gls{iab}, \gls{ncr} represents a low power and cost-effective solution to increase coverage, especially in \gls{mmw}, and serve an ever-expanding network of users, therefore reducing \gls{capex} \cite{green_6g}. 
As defined in \gls{3gpp}'s Release 18, \gls{ncr} is the evolution of the wireless repeater based on the \gls{af} concept, with superior capabilities, such as adaptive \gls{bf} and \gls{tdd}-aware operation \cite{TR_NCR}. 

According to the NextG Alliance, sustainability is a fundamental objective for \gls{6g} and will be considered in the design of networks to reduce CO2 emissions and improve \gls{ee} \cite{nextG}.
In pursuit of environmentally sustainable networks, \gls{nes} has become a vital component, with the aim of reducing the power consumption of the network and increasing \gls{ee}, that is, reducing \gls{opex} for network operators. 
Release 18, commercially recognized as \gls{5g}-Advanced, introduces antenna and transmit power adaptation techniques for the first time to increase \gls{ee}. However, in the release, \gls{nes} and energy-efficient optimization considering repeaters have not been addressed.

This paper advances the current literature by investigating techniques to enable \gls{nes} for repeater-supported wireless networks. In particular, we focus on the optimization of a set of \gls{gnb}'s and \gls{ncr}'s parameters, 
to maximize the network \gls{ee}. Moreover, we investigate trade-offs involving energy-efficient operation and \gls{qos} degradation for the \gls{ue} to quantify the \gls{ee} gains and the corresponding \gls{se} losses. 
To this end, we conduct an analytical study of simple network topologies with and without a repeater (or \gls{ncr}), and 
a detailed system-level simulation study to evaluate the impact of energy-efficient operations based on a realistic network deployment.  
Finally, we provide strategies to enable energy-efficient operations and \gls{nes} based on the availability of different \gls{pa} technologies.

The paper is organized as follows. Section~\ref{sec:related} presents the related work in this domain.
Section~\ref{sec:system} outlines the system model, while Section~\ref{sec:optimization} describes the energy-efficient optimization of the \gls{gnb} and \gls{ncr} configurations. The simulation results are presented and discussed in Section~\ref{sec:results}. Finally, Section~\ref{sec:conclusion} concludes the paper.

\section{Related Work}
\label{sec:related}

In the realm of \gls{ee} and \gls{nes} for future green \gls{6g} networks, \cite{green_6g} provides a survey on new architectural changes associated with \gls{6g} networks with a focus on sustainable networks characterized by ubiquitous coverage, pervasive \gls{ai} and enhanced protocol stack. 
In the context of sustainable networks, the authors of \cite{irs_industry} investigated the performance of \glspl{ris} for \gls{iot}. In particular, they introduced a power consumption model for \gls{ris} to optimize the energy efficiency of the system. In \cite{solar}, the authors considered a cellular network assisted by multiple solar-powered relays. Repeaters are scattered in the cell area so that \glspl{ue} can enjoy higher-quality links and the base station can save energy by reducing its transmit power. They proposed a power control method to reduce interference in the sub-6 GHz frequency range and hence improve network performance. However, they did not consider energy-efficient optimization. Finally, in \cite{powersaving_5g_and_beyond}, the authors explored different potential techniques for saving network power through the adaptation of resources in the time, frequency, spatial, and power domains. In addition, they provide energy efficiency and throughput trade-off analysis to identify which domain has the highest potential for power savings. However, no \gls{ee} optimization was also considered.

To our knowledge, this is the first work to address the energy-efficient optimization of \gls{6g} networks comprising \glspl{sc} and \glspl{ncr}.

\section{System Model}
\label{sec:system}

This section will delve into the critical modeling aspects of our study, laying the foundation for a comprehensive understanding of our results and their implications. We will start by introducing the two specific network topologies for the analytical study. Subsequently, we will introduce the power consumption models, and  
the energy efficiency metrics that serves as a crucial benchmark for evaluating the environmental and operational performance of the network.

\subsection{Direct Topology}

\begin{figure}[t!]
    \centering
    \includegraphics[width=0.6\linewidth]{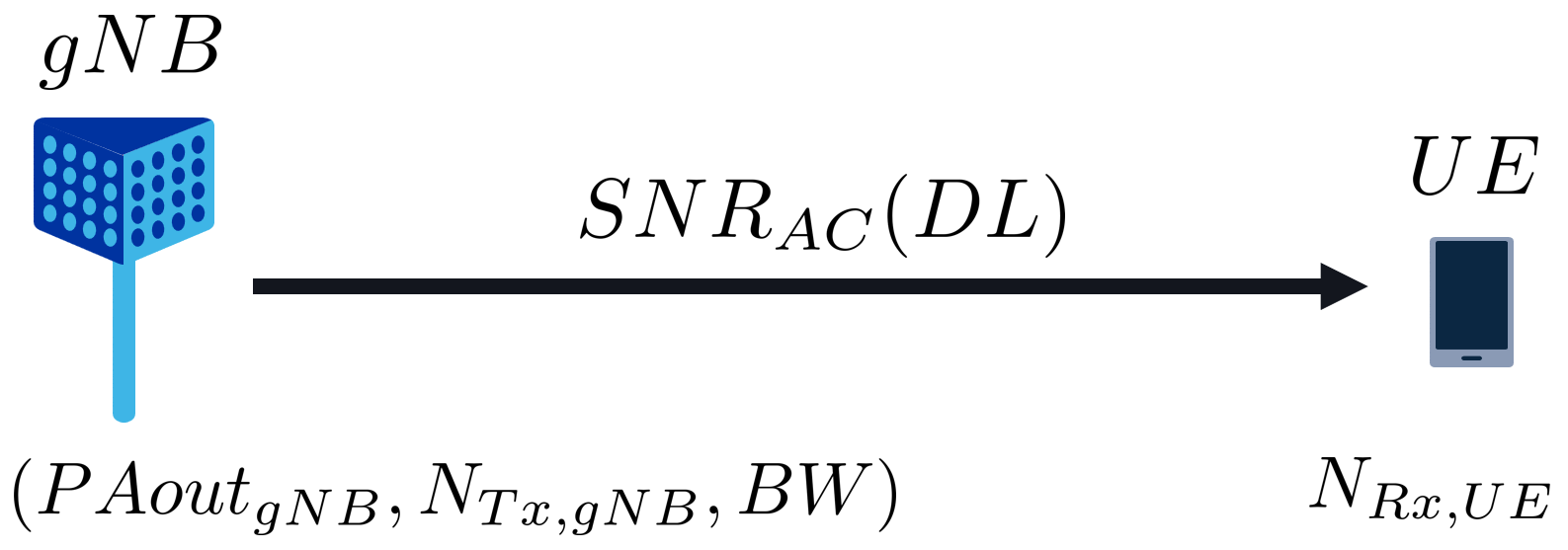}
    \caption{Direct topology. The \gls{ue} is directly served by the \gls{gnb}.}
    \label{fig:simple_top}
\end{figure}

As depicted in Fig.~\ref{fig:simple_top}, the \emph{direct} topology consists of a \gls{ue} directly connected to a \gls{gnb}. To achieve \gls{ee} operations, we assume that we can actively tune the following parameters at the \gls{gnb}: the number of active transmitting antenna elements $N_{Tx,gNB}$, the \gls{pa} output power of each antenna element $PAout_{gNB}$, and the bandwidth $BW$, thus increasing or decreasing its total \gls{eirp}. We consider a fixed number of receiving antennas, $N_{Rx, UE}$, for the \gls{ue}. Given this topology, the expression for \gls{dl} \gls{snr}, $SNR_{AC}(DL)$, can be derived as in Eq.~\ref{eq:direct_snr}, where $PL(d_{AC})$ is the \gls{ac} path loss at the \gls{gnb}-\gls{ue} distance $d_{AC}$, $NF_{UE}$ is the noise figure for the \gls{ue}, and $N_0$ is the thermal noise spectral density. $N_{Tx,gNB}$ is squared since it is considered once to compute the tx power, which is $Tx\ pow = PAout_{gNB} \cdot N_{Tx,gNB}$ in the linear scale, and a second time to include the beamforming gain in the \gls{snr} calculation.

\begin{align}
    SNR_{AC}(DL) = \frac{PAout_{gNB} \cdot N^2_{Tx,gNB} \cdot N_{Rx,UE}}{PL(d_{AC}) \cdot NF_{UE} \cdot N_0 \cdot BW}
    \label{eq:direct_snr}
\end{align}

\subsection{Indirect Topology}

\begin{figure}[t!]
    \centering
    \includegraphics[width=0.58\linewidth]{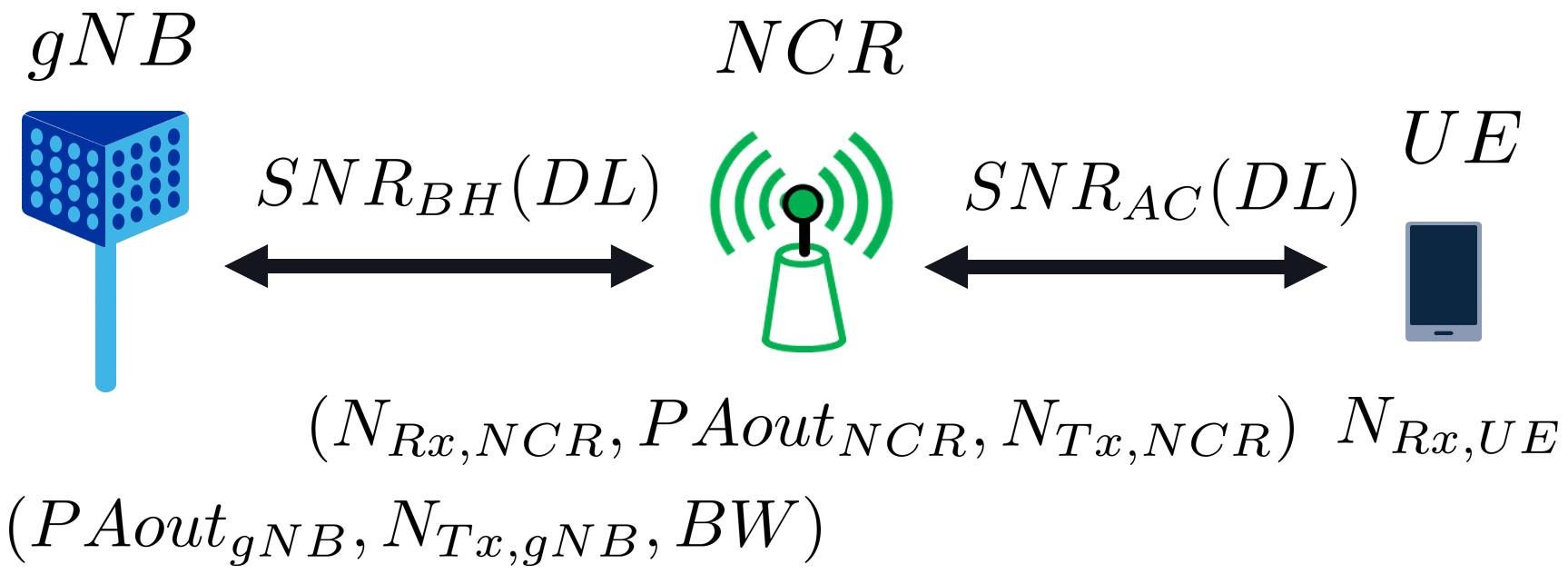}
    \caption{Indirect topology. The \gls{ue} is served by the \gls{gnb} via an \gls{ncr}.}
    \label{fig:repater_top}
\end{figure}

As shown in Fig.~\ref{fig:repater_top}, in the \emph{indirect} topology, a \gls{ue} is connected to a \gls{gnb} through an \gls{ncr}. For \gls{ee} operations, in addition to \gls{gnb}'s parameters, we can also modify the \gls{ncr}'s parameters (i.e., the number of active transmitting and receiving antenna elements, $N_{Tx, NCR}$ and $N_{Rx, NCR}$, and its \gls{pa} output power, $PAout_{NCR}$).


In this topology, we can separate expressions for the \gls{dl} \gls{bh} \gls{snr}, $SNR_{BH}(DL)$, and the \gls{dl} \gls{ac} \gls{snr}, $SNR_{AC}(DL)$, as reported in Eqs.~\ref{eq:bh_snr} and~\ref{eq:ac_snr}, respectively.
$PL(d_{BH})$ and $PL(d_{AC})$ represent the \gls{bh} and \gls{ac} path loss for \gls{gnb}-\gls{ncr} distance $d_{BH}$, and \gls{ncr}-\gls{ue} distance $d_{AC}$. $NF_{NCR}$ is the \gls{ncr}'s noise figure. The actual $PAout_{NCR}$ is the minimum among the maximum value it can achieve given the \gls{ncr}'s  parameters configuration, $PAout_{NCR}^{Max}$, and the $PAout_{NCR}$ calculated based on the signal power received from the \gls{bh} link and the \gls{ncr}'s maximum gain, $G_{max}$, as reported in Eq.~\ref{eq:paout_rep}. $G_{max}$ is assumed to be 90 dB in this study. 

\begin{subequations}
\begin{align}
    \label{eq:bh_snr} SNR_{BH}(DL) &= \frac{PAout_{gNB} \cdot N^2_{Tx,gNB} \cdot N_{Rx,NCR}}{PL(d_{BH}) \cdot NF_{NCR} \cdot N_0 \cdot BW}  \\
    \label{eq:ac_snr} SNR_{AC}(DL) &= \frac{PAout_{NCR} \cdot N^2_{Tx,NCR} \cdot N_{Rx,UE}}{PL(d_{AC}) \cdot NF_{UE} \cdot N_0 \cdot BW} \\
    \label{eq:paout_rep} PAout_{NCR} &= \min \Big\{PAout_{NCR}^{Max}, \nonumber \\
    & \hspace{-10mm} \frac{G_{max} \cdot (1+SNR_{BH}) \cdot NF_{NCR} \cdot N_0 \cdot BW}{N_{Tx,NCR}} \Big\}
\end{align}
\end{subequations}

\noindent
Finally, the effective \gls{dl} \gls{snr}, $effSNR(DL)$, for a \gls{ue} connected to a \gls{gnb} via an \gls{ncr} follows Eq.~\ref{eq:indirect_snr}.

\begin{align}
    effSNR(DL) &= \frac{1}{\frac{1}{SNR_{BH}} + \frac{1}{SNR_{AC}}\cdot \frac{1 + SNR_{BH}}{SNR_{BH}}}
    \label{eq:indirect_snr}
\end{align}

\noindent A through derivation and explanation of Eqs.~\ref{eq:paout_rep} and \ref{eq:indirect_snr} can be found in \cite{Jane}.

\subsection{Power Consumption Model}

Importantly, we model the power consumption of \gls{gnb} and \gls{ncr} based on \gls{3gpp} \cite{TR_NES}, with the details as follows.

\subsubsection{\gls{gnb} power consumption}

\begin{align}
    P_{gNB}(DL) &= P_{ms,gNB} + \alpha \cdot P_{nonPA} + \beta \cdot P_{PA}
    \label{eq:gnb_all}
\end{align}

The expression of the \gls{gnb}'s \gls{dl} power consumption, $P_{gNB}(DL)$, can be divided into three terms, as shown in Eq.~\ref{eq:gnb_all}. $P_{ms,gNB}$ is a fixed power consumption term that does not scale with the \gls{gnb} configuration. $P_{nonPA}$ represents the non \gls{pa}-related power consumption that scales with the number of \glspl{ru}. $P_{PA}$ is the \gls{pa}-related power consumption 
as follows:
\begin{align}
    P_{PA} &= \frac{N_{Tx,gNB} \cdot PAout_{gNB}}{RefTxPower_{perRU}} \cdot \frac{1}{\eta} \cdot (P_{active,DL} - P_{ms,gNB})
\end{align}
where $P_{active,DL}$ and $RefTxPower_{perRU}$ are reference valuesand $\eta$ represents the normalized \gls{pa} efficiency, as in \cite{TR_NES}.
We consider two different cases for the \gls{pa} efficiency ($PA_{eff}$). In the case of \emph{fixed} \gls{pa} efficiency, $\eta$ equals 1,  and the \gls{pa} power consumption increases linearly with the \gls{pa} output power. In practical systems, this assumes that the \gls{pa} exploits bias adjustment to adjust its operational point based on the required output power. This corresponds to a more advanced \gls{pa} technology. In contrast, in the case of \emph{varying} \gls{pa} efficiency,
$\eta = \frac{PAeff(PAout_{gNB})}{PAeff(PAout_{Ref})}$,
where $PAeff(X)$ is the \gls{pa} efficiency at \gls{pa} output power $X$, and $PAout_{Ref}$ is provided by \gls{3gpp}. The \gls{pa} efficiency typically increases with the output power, whose exact function depends on the \gls{pa} implementation. With a varying efficiency model, the \gls{pa}-related power consumption increases slowly with the \gls{pa} output power, and may remain almost constant in some regions. The two constants $\alpha$ and $\beta$ are used to weight the contribution of $P_{nonPA}$ and $P_{PA}$ in the total \gls{gnb}'s power consumption, respectively. In the following, $\alpha = 0.4$ and $\beta = 0.6$, as in \cite{TR_NES}.

\subsubsection{\gls{ncr} power consumption}

We propose a power consumption model to compute the \gls{ncr} power consumption, $P_{NCR}(DL)$, following the essence of the \gls{3gpp} model for \gls{gnb}. It can be expressed as in Eq.~\ref{eq:rep_all}.

\begin{align}
    P_{NCR}(DL) &= P_{const,NCR} +  P_{Rx,NCR} + P_{Tx,NCR}
    \label{eq:rep_all}
\end{align}

$P_{const,NCR}$ is a constant term that also includes the energy consumption associated with the network-controlled functionalities of the \gls{ncr}.
$P_{Rx,NCR}$ represents the power consumption related to the analog receive front-end of \gls{ncr} (including \glspl{lna} and phase shifters) over the \gls{bh} link and can be expressed as: 

\begin{align}
	P_{Rx,NCR} = \frac{N_{Rx,NCR}}{Ref N_{Rx}} \cdot (\gamma \cdot P_{active,UL})
\end{align}
where $Ref N_{Rx}$ and $P_{active,UL}$ are respectively the \gls{3gpp} reference number of receiving antenna elements, and power consumption (of a \gls{gnb}) in the active uplink receive mode \cite{TR_NES}. 
On the other hand, $P_{Tx,NCR}$ is associated with the power consumption of the transmit analogue front-end of \gls{ncr} (primarily \glspl{pa}) over the \gls{ac} link:

\begin{align}
P_{Tx,NCR} = \frac{N_{Tx,NCR} \cdot PAout_{NCR}}{RefTxPower_{perRU}} \cdot \frac{1}{\eta} \cdot (\xi \cdot P_{active,DL})
\end{align}
The two constants $\gamma$ and $\xi$ are used to weight the contribution of \gls{bh} and \gls{ac} configurations in the total \gls{ncr}'s power consumption, as given by $P_{active,UL}$ and $P_{active,DL}$, respectively. In this work, $\gamma = 0.4$ and $\xi = 0.6$. 

The power consumption computed according to the models introduced above is expressed in units of power, referenced as $Unit\ Power$ in the reminder of this paper.






\begin{figure*}[t]
   \centering
    \hspace*{\fill}
    \begin{subfigure}[b]{\FigSize\textwidth}
        \resizebox{\linewidth}{!}{
%
\definecolor{mycolor1}{rgb}{0.00000,0.44700,0.74100}%
\definecolor{mycolor2}{rgb}{0.85000,0.32500,0.09800}%
\begin{tikzpicture}
\begin{scope}
    \clip (0,0) rectangle (6.3,9.05); 
    \shade[left color=green!20, right color=green!20, opacity=0.6] (0,0) rectangle (6.3,9.05);
\end{scope}
\begin{axis}[
    width=4.486in,
    height=3.566in,
    grid=major,
    grid style={dashed},
    scale only axis,
    separate axis lines,
    every outer y axis line/.style={mycolor1},
    every y tick label/.style={font=\color{mycolor1}},
    every y tick/.style={mycolor1},
    ymin=0.998,
    ymax=1.202,
    ytick={1, 1.05, 1.1, 1.15, 1.2},
    yticklabels={100\%, 105\%, 110\%, 115\%, 120\%},
    ylabel style={font=\color{mycolor1}},
    ylabel={Relative EE [\%]},
    yticklabel pos=left,
    ytick pos=left,
    xmin=5,
    xmax=50,
    xtick={5,10,15,20,25,30,35,40,45,50},
    xticklabels={5,10,15,20,25,30,35,40,45,50},
    xlabel style={font=\color{white!15!black}},
    xlabel={gNB-UE distance [m]},
    legend style={at={(0.5,0.2)}, anchor=west}
]
\addplot [color=mycolor1, solid, line width=2.5pt] table[row sep=crcr]{
    1	1.32102851161713\\
2	1.28583451835454\\
3	1.25280106316138\\
4	1.22306635016317\\
5	1.19737410177472\\
6	1.1760661271593\\
7	1.15516806941285\\
8	1.13306431283689\\
9	1.11112009250884\\
10	1.0932797854688\\
11	1.07767864061894\\
12	1.06470111754821\\
13	1.05458161698717\\
14	1.04521561195975\\
15	1.03709218814952\\
16	1.03155051973387\\
17	1.0263378775446\\
18	1.02139957635753\\
19	1.01668684008806\\
20	1.01353955204679\\
21	1.01121476219516\\
22	1.00895580003603\\
23	1.00674934987786\\
24	1.00458434278046\\
25	1.00245161778359\\
26	1.00114256963067\\
27	1.00085202543144\\
28	1.00056340669229\\
29	1.00027606865342\\
30	1\\
31	1\\
32	1\\
33	1\\
34	1\\
35	1\\
36	1\\
37	1\\
38	1\\
39	1\\
40	1\\
41	1\\
42	1\\
43	1\\
44	1\\
45	1\\
46	1\\
47	1\\
48	1\\
49	1\\
50	1\\
51	1\\
52	1\\
53	1\\
54	1\\
55	1\\
56	1\\
57	1\\
58	1\\
59	1\\
60	1\\
61	1\\
62	1\\
63	1\\
64	1\\
65	1\\
66	1\\
67	1\\
68	1\\
69	1\\
70	1\\
71	1\\
72	1\\
73	1\\
74	1\\
75	1\\
76	1\\
77	1\\
78	1\\
79	1\\
80	1\\
81	1\\
82	1\\
83	1\\
84	1\\
85	1\\
86	1\\
87	1\\
88	1\\
89	1\\
90	1\\
91	1\\
92	1\\
93	1\\
94	1\\
95	1\\
96	1\\
97	1\\
98	1\\
99	1\\
100	1\\
101	1\\
102	1\\
103	1\\
104	1\\
105	1\\
106	1\\
107	1\\
108	1\\
109	1\\
110	1\\
111	1\\
112	1\\
113	1\\
114	1\\
115	1\\
116	1\\
117	1\\
118	1\\
119	1\\
120	1\\
121	1\\
122	1\\
123	1\\
124	1\\
125	1\\
126	1\\
127	1\\
128	1\\
129	1\\
130	1\\
131	1\\
132	1\\
133	1\\
134	1\\
135	1\\
136	1\\
137	1\\
138	1\\
139	1\\
140	1\\
141	1\\
142	1\\
143	1\\
144	1\\
145	1\\
146	1\\
147	1\\
148	1\\
149	1\\
150	1\\
151	1\\
152	1\\
153	1\\
154	1\\
155	1\\
156	1\\
157	1\\
158	1\\
159	1\\
160	1\\
161	1\\
162	1\\
163	1\\
164	1\\
165	1\\
166	1\\
167	1\\
168	1\\
169	1\\
170	1\\
171	1\\
172	1\\
173	1\\
174	1\\
175	1\\
176	1\\
177	1\\
178	1\\
179	1\\
180	1\\
181	1\\
182	1\\
183	1\\
184	1\\
185	1\\
186	1\\
187	1\\
188	1\\
189	1\\
190	1\\
191	1\\
192	1\\
193	1\\
194	1\\
195	1\\
196	1\\
197	1\\
198	1\\
199	1\\
200	1\\
201	1\\
202	1\\
203	1\\
204	1\\
205	1\\
206	1\\
207	1\\
208	1\\
209	1\\
210	1\\
211	1\\
212	1\\
213	1\\
214	1\\
215	1\\
216	1\\
217	1\\
218	1\\
219	1\\
220	1\\
221	1\\
222	1\\
223	1\\
224	1\\
225	1\\
226	1\\
227	1\\
228	1\\
229	1\\
230	1\\
231	1\\
232	1\\
233	1\\
234	1\\
235	1\\
236	1\\
237	1\\
238	1\\
239	1\\
240	1\\
241	1\\
242	1\\
243	1\\
244	1\\
245	1\\
246	1\\
247	1\\
248	1\\
249	1\\
250	1\\
251	1\\
252	1\\
253	1\\
254	1\\
255	1\\
256	1\\
257	1\\
258	1\\
259	1\\
260	1\\
261	1\\
262	1\\
263	1\\
264	1\\
265	1\\
266	1\\
267	1\\
268	1\\
269	1\\
270	1\\
271	1\\
272	1\\
273	1\\
274	1\\
275	1\\
276	1\\
277	1\\
278	1\\
279	1\\
280	1\\
281	1\\
282	1\\
283	1\\
284	1\\
285	1\\
286	1\\
287	1\\
288	1\\
289	1\\
290	1\\
291	1\\
292	1\\
293	1\\
294	1\\
295	1\\
296	1\\
297	1\\
298	1\\
299	1\\
300	1\\
};
\addlegendentry{EE (EE-optimal config.)}
\end{axis}

\begin{axis}[
    width=4.486in,
    height=3.566in,
    scale only axis,
    xmin=5,
    xmax=50,
    every outer y axis line/.style={mycolor2},
    every y tick label/.style={font=\color{mycolor2}},
    every y tick/.style={mycolor2},
    ymin=0.899,
    ymax=1.001,
    ytick={0.9, 0.925, 0.95, 0.975, 1},
    yticklabels={90\%, 92.5\%, 95\%, 97.5\%, 100\%},
    ylabel style={font=\color{mycolor2}},
    ylabel={Relative Rate [\%]},
    axis x line*=bottom,
    axis y line*=right,
    xtick={},
    xticklabels={},
    legend style={at={(0.5,0.1)}, anchor=west}
]
\addplot [color=mycolor2, line width=2.5pt]
  table[row sep=crcr]{%
1	0.991964817299867\\
2	0.98264971978956\\
3	0.974077743504041\\
4	0.969269878445272\\
5	0.964843925775995\\
6	0.961368909259826\\
7	0.94428590472684\\
8	0.926217308191965\\
9	0.908279123702042\\
10	0.916299271302397\\
11	0.903223645239051\\
12	0.920059332248571\\
13	0.911314586164038\\
14	0.903220971731463\\
15	0.930184298882622\\
16	0.925213889300187\\
17	0.920538588516361\\
18	0.916109348493288\\
19	0.911882421193358\\
20	0.950870245536444\\
21	0.948689202386646\\
22	0.946569916663097\\
23	0.944499896011779\\
24	0.942468755908625\\
25	0.940467901835057\\
26	0.991232500986834\\
27	0.990944832814507\\
28	0.990659071042587\\
29	0.990374577293608\\
30	1\\
31	1\\
32	1\\
33	1\\
34	1\\
35	1\\
36	1\\
37	1\\
38	1\\
39	1\\
40	1\\
41	1\\
42	1\\
43	1\\
44	1\\
45	1\\
46	1\\
47	1\\
48	1\\
49	1\\
50	1\\
51	1\\
52	1\\
53	1\\
54	1\\
55	1\\
56	1\\
57	1\\
58	1\\
59	1\\
60	1\\
61	1\\
62	1\\
63	1\\
64	1\\
65	1\\
66	1\\
67	1\\
68	1\\
69	1\\
70	1\\
71	1\\
72	1\\
73	1\\
74	1\\
75	1\\
76	1\\
77	1\\
78	1\\
79	1\\
80	1\\
81	1\\
82	1\\
83	1\\
84	1\\
85	1\\
86	1\\
87	1\\
88	1\\
89	1\\
90	1\\
91	1\\
92	1\\
93	1\\
94	1\\
95	1\\
96	1\\
97	1\\
98	1\\
99	1\\
100	1\\
101	1\\
102	1\\
103	1\\
104	1\\
105	1\\
106	1\\
107	1\\
108	1\\
109	1\\
110	1\\
111	1\\
112	1\\
113	1\\
114	1\\
115	1\\
116	1\\
117	1\\
118	1\\
119	1\\
120	1\\
121	1\\
122	1\\
123	1\\
124	1\\
125	1\\
126	1\\
127	1\\
128	1\\
129	1\\
130	1\\
131	1\\
132	1\\
133	1\\
134	1\\
135	1\\
136	1\\
137	1\\
138	1\\
139	1\\
140	1\\
141	1\\
142	1\\
143	1\\
144	1\\
145	1\\
146	1\\
147	1\\
148	1\\
149	1\\
150	1\\
151	1\\
152	1\\
153	1\\
154	1\\
155	1\\
156	1\\
157	1\\
158	1\\
159	1\\
160	1\\
161	1\\
162	1\\
163	1\\
164	1\\
165	1\\
166	1\\
167	1\\
168	1\\
169	1\\
170	1\\
171	1\\
172	1\\
173	1\\
174	1\\
175	1\\
176	1\\
177	1\\
178	1\\
179	1\\
180	1\\
181	1\\
182	1\\
183	1\\
184	1\\
185	1\\
186	1\\
187	1\\
188	1\\
189	1\\
190	1\\
191	1\\
192	1\\
193	1\\
194	1\\
195	1\\
196	1\\
197	1\\
198	1\\
199	1\\
200	1\\
201	1\\
202	1\\
203	1\\
204	1\\
205	1\\
206	1\\
207	1\\
208	1\\
209	1\\
210	1\\
211	1\\
212	1\\
213	1\\
214	1\\
215	1\\
216	1\\
217	1\\
218	1\\
219	1\\
220	1\\
221	1\\
222	1\\
223	1\\
224	1\\
225	1\\
226	1\\
227	1\\
228	1\\
229	1\\
230	1\\
231	1\\
232	1\\
233	1\\
234	1\\
235	1\\
236	1\\
237	1\\
238	1\\
239	1\\
240	1\\
241	1\\
242	1\\
243	1\\
244	1\\
245	1\\
246	1\\
247	1\\
248	1\\
249	1\\
250	1\\
251	1\\
252	1\\
253	1\\
254	1\\
255	1\\
256	1\\
257	1\\
258	1\\
259	1\\
260	1\\
261	1\\
262	1\\
263	1\\
264	1\\
265	1\\
266	1\\
267	1\\
268	1\\
269	1\\
270	1\\
271	1\\
272	1\\
273	1\\
274	1\\
275	1\\
276	1\\
277	1\\
278	1\\
279	1\\
280	1\\
281	1\\
282	1\\
283	1\\
284	1\\
285	1\\
286	1\\
287	1\\
288	1\\
289	1\\
290	1\\
291	1\\
292	1\\
293	1\\
294	1\\
295	1\\
296	1\\
297	1\\
298	1\\
299	1\\
300	1\\
};
\addlegendentry{Rate (EE-optimal config.)}
\end{axis}

\end{tikzpicture}}
        \caption{Relative \gls{ee} and rate for fixed \gls{pa} efficiency.}
        \label{fig:rel_ee_direct_fixed}
    \end{subfigure}%
    \hspace*{\fill}
    \begin{subfigure}[b]{\FigSize\textwidth}
        \resizebox{\linewidth}{!}{
%
\definecolor{mycolor1}{rgb}{0.00000,0.44700,0.74100}%
\definecolor{mycolor2}{rgb}{0.85000,0.32500,0.09800}%
\begin{tikzpicture}
\begin{scope}
    \clip (0,0) rectangle (7.6,9.05); 
    \shade[left color=green!20, right color=green!20, opacity=0.6] (0,0) rectangle (7.6,9.05);
\end{scope}
\begin{axis}[
    width=4.486in,
    height=3.566in,
    grid=major,
    grid style={dashed},
    scale only axis,
    separate axis lines,
    every outer y axis line/.style={mycolor1},
    every y tick label/.style={font=\color{mycolor1}},
    every y tick/.style={mycolor1},
    ymin=0.995,
    ymax=1.605,
    ytick={1, 1.1, 1.2, 1.3, 1.4, 1.5, 1.6},
    yticklabels={100\%, 110\%, 120\%, 130\%, 140\%, 150\%, 160\%},
    ylabel style={font=\color{mycolor1}},
    ylabel={Relative EE [\%]},
    yticklabel pos=left,
    ytick pos=left,
    xmin=5,
    xmax=50,
    xtick={5,10,15,20,25,30,35,40,45,50},
    xticklabels={5,10,15,20,25,30,35,40,45,50},
    xlabel style={font=\color{white!15!black}},
    xlabel={gNB-UE distance [m]},
    legend style={at={(0.5,0.2)}, anchor=west}
]
\addplot [color=mycolor1, solid, line width=2.5pt] table[row sep=crcr]{
    1	2.18882521881999\\
2	1.99763592538211\\
3	1.83651657870902\\
4	1.70147770359614\\
5	1.58841575462503\\
6	1.49388628099314\\
7	1.41585370727114\\
8	1.35109577526845\\
9	1.2976066271161\\
10	1.25305239568767\\
11	1.21628472638611\\
12	1.1853482261126\\
13	1.15916509798578\\
14	1.13708811346516\\
15	1.11813277879506\\
16	1.10172006089171\\
17	1.08741845853818\\
18	1.07498922360078\\
19	1.06409157955267\\
20	1.05446660983713\\
21	1.04594956844389\\
22	1.03840820170416\\
23	1.03182372043877\\
24	1.02602145022862\\
25	1.02092047591111\\
26	1.01646265924678\\
27	1.01261503954933\\
28	1.00938265899322\\
29	1.00666365941959\\
30	1.00442901953668\\
31	1.00265375983359\\
32	1.00137637754748\\
33	1.00052327042799\\
34	1.00006919344948\\
35	1\\
36	1\\
37	1\\
38	1\\
39	1\\
40	1\\
41	1\\
42	1\\
43	1\\
44	1\\
45	1\\
46	1\\
47	1\\
48	1\\
49	1\\
50	1\\
51	1\\
52	1\\
53	1\\
54	1\\
55	1\\
56	1\\
57	1\\
58	1\\
59	1\\
60	1\\
61	1\\
62	1\\
63	1\\
64	1\\
65	1\\
66	1\\
67	1\\
68	1\\
69	1\\
70	1\\
71	1\\
72	1\\
73	1\\
74	1\\
75	1\\
76	1\\
77	1\\
78	1\\
79	1\\
80	1\\
81	1\\
82	1\\
83	1\\
84	1\\
85	1\\
86	1\\
87	1\\
88	1\\
89	1\\
90	1\\
91	1\\
92	1\\
93	1\\
94	1\\
95	1\\
96	1\\
97	1\\
98	1\\
99	1\\
100	1\\
101	1\\
102	1\\
103	1\\
104	1\\
105	1\\
106	1\\
107	1\\
108	1\\
109	1\\
110	1\\
111	1\\
112	1\\
113	1\\
114	1\\
115	1\\
116	1\\
117	1\\
118	1\\
119	1\\
120	1\\
121	1\\
122	1\\
123	1\\
124	1\\
125	1\\
126	1\\
127	1\\
128	1\\
129	1\\
130	1\\
131	1\\
132	1\\
133	1\\
134	1\\
135	1\\
136	1\\
137	1\\
138	1\\
139	1\\
140	1\\
141	1\\
142	1\\
143	1\\
144	1\\
145	1\\
146	1\\
147	1\\
148	1\\
149	1\\
150	1\\
151	1\\
152	1\\
153	1\\
154	1\\
155	1\\
156	1\\
157	1\\
158	1\\
159	1\\
160	1\\
161	1\\
162	1\\
163	1\\
164	1\\
165	1\\
166	1\\
167	1\\
168	1\\
169	1\\
170	1\\
171	1\\
172	1\\
173	1\\
174	1\\
175	1\\
176	1\\
177	1\\
178	1\\
179	1\\
180	1\\
181	1\\
182	1\\
183	1\\
184	1\\
185	1\\
186	1\\
187	1\\
188	1\\
189	1\\
190	1\\
191	1\\
192	1\\
193	1\\
194	1\\
195	1\\
196	1\\
197	1\\
198	1\\
199	1\\
200	1\\
201	1\\
202	1\\
203	1\\
204	1\\
205	1\\
206	1\\
207	1\\
208	1\\
209	1\\
210	1\\
211	1\\
212	1\\
213	1\\
214	1\\
215	1\\
216	1\\
217	1\\
218	1\\
219	1\\
220	1\\
221	1\\
222	1\\
223	1\\
224	1\\
225	1\\
226	1\\
227	1\\
228	1\\
229	1\\
230	1\\
231	1\\
232	1\\
233	1\\
234	1\\
235	1\\
236	1\\
237	1\\
238	1\\
239	1\\
240	1\\
241	1\\
242	1\\
243	1\\
244	1\\
245	1\\
246	1\\
247	1\\
248	1\\
249	1\\
250	1\\
251	1\\
252	1\\
253	1\\
254	1\\
255	1\\
256	1\\
257	1\\
258	1\\
259	1\\
260	1\\
261	1\\
262	1\\
263	1\\
264	1\\
265	1\\
266	1\\
267	1\\
268	1\\
269	1\\
270	1\\
271	1\\
272	1\\
273	1\\
274	1\\
275	1\\
276	1\\
277	1\\
278	1\\
279	1\\
280	1\\
281	1\\
282	1\\
283	1\\
284	1\\
285	1\\
286	1\\
287	1\\
288	1\\
289	1\\
290	1\\
291	1\\
292	1\\
293	1\\
294	1\\
295	1\\
296	1\\
297	1\\
298	1\\
299	1\\
300	1\\
};
\addlegendentry{EE (EE-optimal config.)}
\end{axis}

\begin{axis}[
    width=4.486in,
    height=3.566in,
    scale only axis,
    xmin=5,
    xmax=50,
    every outer y axis line/.style={mycolor2},
    every y tick label/.style={font=\color{mycolor2}},
    every y tick/.style={mycolor2},
    ymin=0.698, 
    ymax=1.0017,
    ytick={0.7, 0.75, 0.8, 0.85, 0.9, 0.95, 1},
    yticklabels={70\%, 75\%, 80\%, 85\%, 90\%, 95\%, 100\%},
    ylabel style={font=\color{mycolor2}},
    ylabel={Relative Rate [\%]},
    axis x line*=bottom,
    axis y line*=right,
    xtick={},
    xticklabels={},
    legend style={at={(0.5,0.1)}, anchor=west}
]
\addplot [color=mycolor2, line width=2.5pt]
  table[row sep=crcr]{%
1	0.964297226612554\\
2	0.928657800527268\\
3	0.898427790488667\\
4	0.873752735860672\\
5	0.854328668061832\\
6	0.839822890088957\\
7	0.813174472050671\\
8	0.808845379440128\\
9	0.792604932980266\\
10	0.795869169777938\\
11	0.787308669580591\\
12	0.781699313112764\\
13	0.778529973217951\\
14	0.791360609939449\\
15	0.791767143925081\\
16	0.793544003128247\\
17	0.796467907947152\\
18	0.813511973989951\\
19	0.818206368078343\\
20	0.823629759937554\\
21	0.8296978992863\\
22	0.836344691042765\\
23	0.856139261649615\\
24	0.863803244108602\\
25	0.871925041025663\\
26	0.880479879093943\\
27	0.90177754895937\\
28	0.911174939070464\\
29	0.920963381026766\\
30	0.931134702247235\\
31	0.94168311434305\\
32	0.964840596265504\\
33	0.976186833635927\\
34	0.987906497471568\\
35	1\\
36	1\\
37	1\\
38	1\\
39	1\\
40	1\\
41	1\\
42	1\\
43	1\\
44	1\\
45	1\\
46	1\\
47	1\\
48	1\\
49	1\\
50	1\\
51	1\\
52	1\\
53	1\\
54	1\\
55	1\\
56	1\\
57	1\\
58	1\\
59	1\\
60	1\\
61	1\\
62	1\\
63	1\\
64	1\\
65	1\\
66	1\\
67	1\\
68	1\\
69	1\\
70	1\\
71	1\\
72	1\\
73	1\\
74	1\\
75	1\\
76	1\\
77	1\\
78	1\\
79	1\\
80	1\\
81	1\\
82	1\\
83	1\\
84	1\\
85	1\\
86	1\\
87	1\\
88	1\\
89	1\\
90	1\\
91	1\\
92	1\\
93	1\\
94	1\\
95	1\\
96	1\\
97	1\\
98	1\\
99	1\\
100	1\\
101	1\\
102	1\\
103	1\\
104	1\\
105	1\\
106	1\\
107	1\\
108	1\\
109	1\\
110	1\\
111	1\\
112	1\\
113	1\\
114	1\\
115	1\\
116	1\\
117	1\\
118	1\\
119	1\\
120	1\\
121	1\\
122	1\\
123	1\\
124	1\\
125	1\\
126	1\\
127	1\\
128	1\\
129	1\\
130	1\\
131	1\\
132	1\\
133	1\\
134	1\\
135	1\\
136	1\\
137	1\\
138	1\\
139	1\\
140	1\\
141	1\\
142	1\\
143	1\\
144	1\\
145	1\\
146	1\\
147	1\\
148	1\\
149	1\\
150	1\\
151	1\\
152	1\\
153	1\\
154	1\\
155	1\\
156	1\\
157	1\\
158	1\\
159	1\\
160	1\\
161	1\\
162	1\\
163	1\\
164	1\\
165	1\\
166	1\\
167	1\\
168	1\\
169	1\\
170	1\\
171	1\\
172	1\\
173	1\\
174	1\\
175	1\\
176	1\\
177	1\\
178	1\\
179	1\\
180	1\\
181	1\\
182	1\\
183	1\\
184	1\\
185	1\\
186	1\\
187	1\\
188	1\\
189	1\\
190	1\\
191	1\\
192	1\\
193	1\\
194	1\\
195	1\\
196	1\\
197	1\\
198	1\\
199	1\\
200	1\\
201	1\\
202	1\\
203	1\\
204	1\\
205	1\\
206	1\\
207	1\\
208	1\\
209	1\\
210	1\\
211	1\\
212	1\\
213	1\\
214	1\\
215	1\\
216	1\\
217	1\\
218	1\\
219	1\\
220	1\\
221	1\\
222	1\\
223	1\\
224	1\\
225	1\\
226	1\\
227	1\\
228	1\\
229	1\\
230	1\\
231	1\\
232	1\\
233	1\\
234	1\\
235	1\\
236	1\\
237	1\\
238	1\\
239	1\\
240	1\\
241	1\\
242	1\\
243	1\\
244	1\\
245	1\\
246	1\\
247	1\\
248	1\\
249	1\\
250	1\\
251	1\\
252	1\\
253	1\\
254	1\\
255	1\\
256	1\\
257	1\\
258	1\\
259	1\\
260	1\\
261	1\\
262	1\\
263	1\\
264	1\\
265	1\\
266	1\\
267	1\\
268	1\\
269	1\\
270	1\\
271	1\\
272	1\\
273	1\\
274	1\\
275	1\\
276	1\\
277	1\\
278	1\\
279	1\\
280	1\\
281	1\\
282	1\\
283	1\\
284	1\\
285	1\\
286	1\\
287	1\\
288	1\\
289	1\\
290	1\\
291	1\\
292	1\\
293	1\\
294	1\\
295	1\\
296	1\\
297	1\\
298	1\\
299	1\\
300	1\\
};
\addlegendentry{Rate (EE-optimal config.)}
\end{axis}

\end{tikzpicture}}
        \caption{Relative \gls{ee} and rate for varying \gls{pa} efficiency.}
        \label{fig:rel_ee_direct_varying}
    \end{subfigure}
    \hspace*{\fill}
    \caption{Link-level simulation results for the direct topology. Relative \gls{ee} and rate are computed for the \gls{ee}-optimal \gls{gnb} configuration (resulting from the optimization in (\ref{opt:direct})) against the baseline configuration (i.e., the most spectral-efficient \gls{gnb} configuration) given increasing \gls{gnb}-\gls{ue} distance. 
    }
    \label{fig:direct_overall}
\end{figure*}
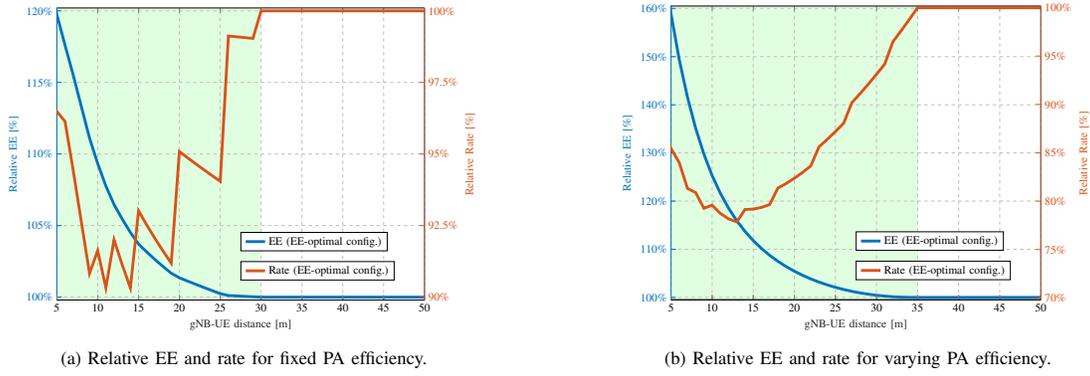

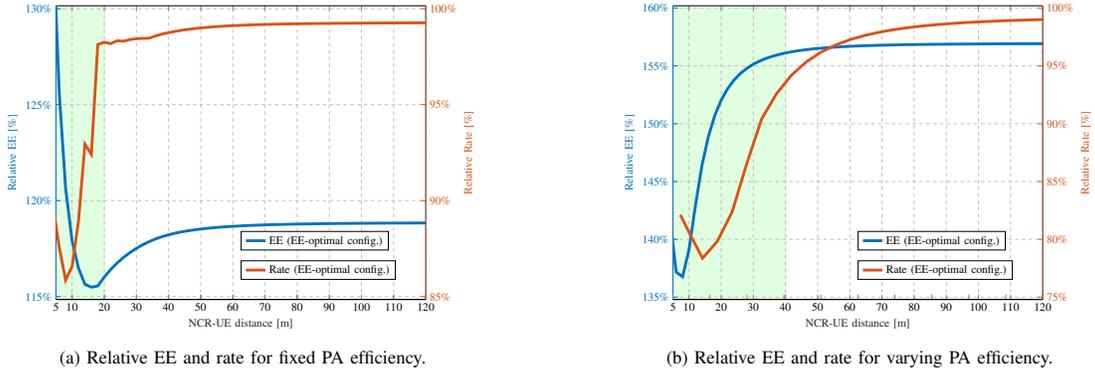
\begin{figure*}[t]
    \centering
    \hspace*{\fill}%
    \begin{subfigure}[b]{\FigSize\textwidth}
        \centering
        \resizebox{\linewidth}{!}{
%
\definecolor{mycolor1}{rgb}{0.00000,0.44700,0.74100}%
\definecolor{mycolor2}{rgb}{0.85000,0.32500,0.09800}%
\begin{tikzpicture}
\begin{scope}
    \clip (0,0) rectangle (1.5,9.05); 
    \shade[left color=green!20, right color=green!20, opacity=0.6] (0,0) rectangle (1.5,9.05);
\end{scope}
\begin{axis}[
    width=4.486in,
    height=3.566in,
    grid=major,
    grid style={dashed},
    scale only axis,
    separate axis lines,
    every outer y axis line/.style={mycolor1},
    every y tick label/.style={font=\color{mycolor1}},
    every y tick/.style={mycolor1},
    ymin=1.1485,
    ymax=1.3015,
    ytick={1.15, 1.2, 1.25, 1.3},
    yticklabels={115\%, 120\%, 125\%, 130\%},
    ylabel style={font=\color{mycolor1}},
    ylabel={Relative EE [\%]},
    yticklabel pos=left,
    ytick pos=left,
    xmin=5,
    xmax=120,
    xtick={5,10,20,30,40,50,60,70,80,90,100,110,120},
    xticklabels={5,10,20,30,40,50,60,70,80,90,100,110,120},
    xlabel style={font=\color{white!15!black}},
    xlabel={NCR-UE distance [m]},
    legend style={at={(0.5,0.2)}, anchor=west}
]
\addplot [color=mycolor1, solid, line width=2.5pt] table[row sep=crcr]{
    1	1.49500080481067\\
2	1.45147143477749\\
4	1.34045516775106\\
6	1.25644594540722\\
8	1.20623746866011\\
10	1.1790335692799\\
12	1.16478075619627\\
14	1.1564004247201\\
16	1.15489736028002\\
18	1.15552922128533\\
20	1.1601446052687\\
22	1.16409514800898\\
24	1.1674730029835\\
26	1.17041101845076\\
28	1.17291677542746\\
30	1.17511636217531\\
32	1.17701023384957\\
34	1.17865621771536\\
36	1.18005932777937\\
38	1.18121895463704\\
40	1.18219065974149\\
42	1.18300067951803\\
44	1.18368648856362\\
46	1.18427194911698\\
48	1.18475929749669\\
50	1.18518831676275\\
52	1.18555507877217\\
54	1.18587191226269\\
56	1.18615409504994\\
58	1.18639389109961\\
60	1.18660439629177\\
62	1.18678600352757\\
64	1.18695536453356\\
66	1.18709551899977\\
68	1.18722322368849\\
70	1.18733670321323\\
72	1.18744814774828\\
74	1.18753412709928\\
76	1.18762169663544\\
78	1.18769562664483\\
80	1.18775411271118\\
82	1.18781926471203\\
84	1.18787476042908\\
86	1.18792045453767\\
88	1.18797030043312\\
90	1.18800752452899\\
92	1.18805065291881\\
94	1.18807994297772\\
96	1.18811255657416\\
98	1.18814781583166\\
100	1.1881786725674\\
102	1.18819320055654\\
104	1.18821950235226\\
106	1.18824709841353\\
108	1.1882610748922\\
110	1.18827411639927\\
112	1.18829750477882\\
114	1.1883069855242\\
116	1.18831805621532\\
118	1.18833665004991\\
120	1.18834697258014\\
122	1.18836359671192\\
124	1.18837458729738\\
126	1.18838183194871\\
128	1.18839803081409\\
130	1.18840454985232\\
132	1.18841451299824\\
134	1.18841861639405\\
136	1.18842110643369\\
138	1.18843145692337\\
140	1.18843168099302\\
142	1.18844923079333\\
144	1.188456521171\\
146	1.18846012517967\\
148	1.18845714477957\\
150	1.18846655698001\\
152	1.18846596896799\\
154	1.18847581783871\\
156	1.1884774531235\\
158	1.18848734978045\\
160	1.1884921413474\\
162	1.18848929978711\\
164	1.1884983438795\\
};
\addlegendentry{EE (EE-optimal config.)}
\end{axis}

\begin{axis}[
    width=4.486in,
    height=3.566in,
    scale only axis,
    xmin=5,
    xmax=120,
    every outer y axis line/.style={mycolor2},
    every y tick label/.style={font=\color{mycolor2}},
    every y tick/.style={mycolor2},
    ymin=0.8485,
    ymax=1.0015,
    ytick={0.85, 0.90, 0.95, 1},
    yticklabels={85\%, 90\%, 95\%, 100\%},
    ylabel style={font=\color{mycolor2}},
    ylabel={Relative Rate [\%]},
    axis x line*=bottom,
    axis y line*=right,
    xtick={},
    xticklabels={},
    legend style={at={(0.5,0.1)}, anchor=west}
]
\addplot [color=mycolor2, line width=2.5pt]
  table[row sep=crcr]{%
1	0.94002037734473\\
2	0.922022906328756\\
4	0.897803395582972\\
6	0.875669565731893\\
8	0.85850817181222\\
10	0.865708382391655\\
12	0.889355511241537\\
14	0.929413795747306\\
16	0.924083910205987\\
18	0.981309863023824\\
20	0.982469574370613\\
22	0.981782698249637\\
24	0.983241653358425\\
26	0.983054239427195\\
28	0.983887793272548\\
30	0.984336365898051\\
32	0.984523956220472\\
34	0.984620384853874\\
36	0.985793282875199\\
38	0.986759903426601\\
40	0.987447987360863\\
42	0.988126752030895\\
44	0.988699019339964\\
46	0.989184518102718\\
48	0.989598768309235\\
50	0.989954096665465\\
52	0.990260379642443\\
54	0.990525593938307\\
56	0.990756229076539\\
58	0.990957600088669\\
60	0.991134086921531\\
62	0.991289319495996\\
64	0.991426322022027\\
66	0.991547626460339\\
68	0.991655362399089\\
70	0.991751328742626\\
72	0.991837051259786\\
74	0.991913829055039\\
76	0.991982772301239\\
78	0.992044833034607\\
80	0.9921008304077\\
82	0.992151471491306\\
84	0.992197368482221\\
86	0.9922390529953\\
88	0.992276987978697\\
90	0.992311577683861\\
92	0.992343176036699\\
94	0.992372093689426\\
96	0.992398603980344\\
98	0.992422947985654\\
100	0.992445338814504\\
102	0.992465965271062\\
104	0.99248499498539\\
106	0.992502577097264\\
108	0.992518844562761\\
110	0.992533916141693\\
112	0.992547898113996\\
114	0.992560885765556\\
116	0.992572964677538\\
118	0.992584211847558\\
120	0.99259469666671\\
122	0.992604481772795\\
124	0.992613623797046\\
126	0.992622174018777\\
128	0.992630178940562\\
130	0.992637680794376\\
132	0.992644717988102\\
134	0.992651325499628\\
136	0.992657535225952\\
138	0.99266337629225\\
140	0.992668875326607\\
142	0.992674056704054\\
144	0.992678942764044\\
146	0.992683554004557\\
148	0.992687909255221\\
150	0.992692025832494\\
152	0.992695919678504\\
154	0.992699605485661\\
156	0.992703096808686\\
158	0.992706406165244\\
160	0.992709545126747\\
162	0.992712524400138\\
164	0.992715353901792\\
};
\addlegendentry{Rate (EE-optimal config.)}
\end{axis}

\end{tikzpicture}}
        \caption{Relative \gls{ee} and rate for fixed \gls{pa} efficiency.}
        \label{fig:ee_snr_indirect_fixed}
    \end{subfigure}%
    \hspace*{\fill}
    \begin{subfigure}[b]{\FigSize\textwidth}
        \centering
        \resizebox{\linewidth}{!}{
%
\definecolor{mycolor1}{rgb}{0.00000,0.44700,0.74100}%
\definecolor{mycolor2}{rgb}{0.85000,0.32500,0.09800}%
\begin{tikzpicture}
\begin{scope}
    \clip (0,0) rectangle (3.5,9.05); 
    \shade[left color=green!20, right color=green!20, opacity=0.6] (0,0) rectangle (3.5,9.05);
\end{scope}
\begin{axis}[
    width=4.486in,
    height=3.566in,
    grid=major,
    grid style={dashed},
    scale only axis,
    separate axis lines,
    every outer y axis line/.style={mycolor1},
    every y tick label/.style={font=\color{mycolor1}},
    every y tick/.style={mycolor1},
    ymin=1.348,
    ymax=1.602,
    ytick={1.35, 1.4, 1.45, 1.5, 1.55, 1.6},
    yticklabels={135\%, 140\%, 145\%, 150\%, 155\%, 160\%},
    ylabel style={font=\color{mycolor1}},
    ylabel={Relative EE [\%]},
    yticklabel pos=left,
    ytick pos=left,
    xmin=5,
    xmax=120,
    xtick={5,10,20,30,40,50,60,70,80,90,100,110,120},
    xticklabels={5,10,20,30,40,50,60,70,80,90,100,110,120},
    xlabel style={font=\color{white!15!black}},
    xlabel={NCR-UE distance [m]},
    legend style={at={(0.5,0.2)}, anchor=west}
]
\addplot [color=mycolor1, solid, line width=2.5pt] table[row sep=crcr]{
   1	1.58363463598987\\
2	1.52110219432908\\
4	1.4203206344299\\
6	1.37173929833529\\
8	1.36758891581672\\
10	1.39123936974223\\
12	1.42932640974628\\
14	1.4635896724332\\
16	1.48871384198444\\
18	1.50698417201131\\
20	1.52035768304456\\
22	1.53027648994395\\
24	1.53774894079323\\
26	1.5434746810668\\
28	1.54792401022864\\
30	1.55143663690687\\
32	1.55423665953092\\
34	1.55650132742704\\
36	1.55835079874237\\
38	1.55987779465724\\
40	1.56114495217495\\
42	1.56220240546401\\
44	1.56310115178443\\
46	1.56385874432873\\
48	1.5645003706211\\
50	1.5650640684687\\
52	1.56553644148048\\
54	1.56595018326582\\
56	1.566318456735\\
58	1.56662892102338\\
60	1.56690936757581\\
62	1.56714270689778\\
64	1.56735974039887\\
66	1.567546440538\\
68	1.56771664393531\\
70	1.56786812763408\\
72	1.56800704255859\\
74	1.56812150562042\\
76	1.56822894186237\\
78	1.56832771493381\\
80	1.56841613580182\\
82	1.56849179625046\\
84	1.56856696091357\\
86	1.56862701858678\\
88	1.56868394550081\\
90	1.56874497349215\\
92	1.56879066285985\\
94	1.56883025665037\\
96	1.56887366706271\\
98	1.56891950835991\\
100	1.56896047971415\\
102	1.56898087550497\\
104	1.56901625476966\\
106	1.5690415680154\\
108	1.56907265135569\\
110	1.56908851245783\\
112	1.56911186951068\\
114	1.56913198606687\\
116	1.56914846520172\\
118	1.56916187318254\\
120	1.56918972751826\\
122	1.56919779878878\\
124	1.56922312765898\\
126	1.56923261122001\\
128	1.56924450506644\\
130	1.5692553566658\\
132	1.56926620755842\\
134	1.56927312812522\\
136	1.56928310168899\\
138	1.56928716252579\\
140	1.56930504631278\\
142	1.5693171693138\\
144	1.56932087509984\\
146	1.5693379327336\\
148	1.56932700890151\\
150	1.56933839669515\\
152	1.56934349655349\\
154	1.56935495303488\\
156	1.56935487626254\\
158	1.56936503064504\\
160	1.56936831479406\\
162	1.56937007656258\\
164	1.56937819207803\\
};
\addlegendentry{EE (EE-optimal config.)}
\end{axis}

\begin{axis}[
    width=4.486in,
    height=3.566in,
    scale only axis,
    xmin=0,
    xmax=50,
    every outer y axis line/.style={mycolor2},
    every y tick label/.style={font=\color{mycolor2}},
    every y tick/.style={mycolor2},
    ymin=0.748,
    ymax=1.002,
    ytick={0.75, 0.80, 0.85, 0.9, 0.95, 1},
    yticklabels={75\%, 80\%, 85\%, 90\%, 95\%, 100\%},
    ylabel style={font=\color{mycolor2}},
    ylabel={Relative Rate [\%]},
    axis x line*=bottom,
    axis y line*=right,
    xtick={},
    xticklabels={},
    legend style={at={(0.5,0.1)}, anchor=west}
]
\addplot [color=mycolor2, line width=2.5pt]
  table[row sep=crcr]{%
1	0.821215964904309\\
2	0.808351610401169\\
4	0.783618432667828\\
6	0.798275425705621\\
8	0.823615845911506\\
10	0.865890843018193\\
12	0.904203451253451\\
14	0.925878578810487\\
16	0.941773434077495\\
18	0.953331146303524\\
20	0.961790304756436\\
22	0.968064765904435\\
24	0.972792875425897\\
26	0.976413468376423\\
28	0.979228767811703\\
30	0.981449044149973\\
32	0.98322265338425\\
34	0.984655896805379\\
36	0.985826144260576\\
38	0.986790571809534\\
40	0.987592046980407\\
42	0.988263146374068\\
44	0.98882893744332\\
46	0.989308932310771\\
48	0.989718479489301\\
50	0.990069768817889\\
52	0.990372566654079\\
54	0.990634760454441\\
56	0.990862766925514\\
58	0.991061841311096\\
60	0.991236314183118\\
62	0.991389774466959\\
64	0.991525212163323\\
66	0.991645130552186\\
68	0.991751635069624\\
70	0.991846504196673\\
72	0.991931246363716\\
74	0.992007145900339\\
76	0.992075300344179\\
78	0.992136650888971\\
80	0.992192007352858\\
82	0.992242068745203\\
84	0.992287440279913\\
86	0.992328647505533\\
88	0.992366148085539\\
90	0.992400341655157\\
92	0.992431578097525\\
94	0.992460164515432\\
96	0.992486371123275\\
98	0.992510436241431\\
100	0.992532570542452\\
102	0.99255296067136\\
104	0.992571772340881\\
106	0.992589152984661\\
108	0.9926052340376\\
110	0.992620132900525\\
112	0.99263395463706\\
114	0.99264679344246\\
116	0.992658733918267\\
118	0.992669852180613\\
120	0.992680216826091\\
122	0.992689889775321\\
124	0.992698927011028\\
126	0.992707379225274\\
128	0.992715292388059\\
130	0.992722708247654\\
132	0.992729664772004\\
134	0.992736196538308\\
136	0.992742335078195\\
138	0.992748109183183\\
140	0.992753545176273\\
142	0.992758667153232\\
144	0.992763497197563\\
146	0.992768055572377\\
148	0.992772360891551\\
150	0.992776430273089\\
152	0.99278027947646\\
154	0.992783923025701\\
156	0.992787374320183\\
158	0.992790645734109\\
160	0.992793748706308\\
162	0.992796693821004\\
164	0.99279949088088\\
};
\addlegendentry{Rate (EE-optimal config.)}
\end{axis}

\end{tikzpicture}}
        \caption{Relative \gls{ee} and rate for varying \gls{pa} efficiency.}
        \label{fig:ee_snr_indirect_varying}
    \end{subfigure}
    \hspace*{\fill}%
    \caption{Link-level simulation results for the indirect topology. Relative \gls{ee} and rate for the \gls{ee}-optimal configuration are obtained after optimization of (\ref{opt:indirect}) for each \gls{ncr}-\gls{ue} distance against the baseline case, where both \gls{gnb} and \gls{ncr} have their most spectral-efficient configuration.}
    \label{fig:indirect_overall}
\end{figure*}

\section{Optimization}
\label{sec:optimization}

Our objective is to maximize \gls{ee} given the system model described in the previous section. We adopt the same definition of \gls{ee} as in \cite{etsi}.
For the \emph{direct} topology, we can express the \gls{dl} \gls{ee} as in Eq.~\ref{eq:ee_direct}. The numerator is Shannon's rate at $SNR_{AC}(DL)$, while the denominator represents the corresponding total \gls{dl} network power consumption given by the \gls{gnb}. 
For the \emph{indirect} topology, instead, the \gls{dl} \gls{ee} can be defined as in Eq.~\ref{eq:ee_indirect}. In this case, the total \gls{dl} network power consumption is the sum of \gls{gnb}'s and \gls{ncr}'s power consumption. 
In both cases, \gls{ee} has units of $Bps$ over $Unit\ Power$.

\begin{equation}
\begin{aligned}
    EE_{direct}(DL) = \frac{Rate[SNR_{AC}(DL)]}{P_{Total}(DL)} = \\ = \frac{BW \cdot log_2[1 + SNR_{AC}(DL)]}{P_{gNB}(DL)}
    \label{eq:ee_direct}
\end{aligned}
\end{equation}

\begin{equation}
\begin{aligned}
    EE_{indirect}(DL) = \frac{Rate[effSNR(DL)]}{P_{Total}(DL)} = \\ = \frac{BW \cdot log_2[1 + effSNR(DL)]}{P_{gNB}(DL) + P_{NCR}(DL)}
    \label{eq:ee_indirect}
\end{aligned}
\end{equation}

We can define the \gls{ee}-optimal optimization problem for the \emph{direct} topology as follows:

\begin{maxi}|l|[3]
	{} {EE_{direct}(DL) \label{opt:direct}}
	{}{}
	\addConstraint{BW \in \interval{1}{400}\ MHz}
	\addConstraint{PAout_{gNB} \in \interval{0}{10}\ dBm}
	\addConstraint{N_{Tx,gNB} \in \interval{1}{192}}
\end{maxi}
where optimization is over $BW$, $PAout_{gNB}$, and $N_{Tx, gNB}$. The ranges of values were determined on the basis of our study. Similarly, the \gls{ee} maximization problem for the \emph{indirect} topology can be expressed as follows:


\begin{maxi}|l|[3]
	{} {EE_{indirect}(DL) \label{opt:indirect}}
	{}{}
	\addConstraint{BW \in \interval{1}{400}\ MHz}
	\addConstraint{PAout_{gNB} \in \interval{0}{10}\ dBm}
	\addConstraint{N_{Tx,gNB} \in \interval{1}{192}}
	\addConstraint{PAout_{NCR} \in \interval{0}{10}\ dBm}
	\addConstraint{N_{Tx,NCR} \in \interval{1}{32}},{\ N_{Rx,NCR} \in \interval{1}{32}}
\end{maxi}
where optimization is for $BW$, $PAout_{gNB}$, $PAout_{NCR}$, $N_{Tx,gNB}$, $N_{Tx,NCR}$, and $N_{Rx,NCR}$.


Optimization problems \ref{opt:direct} and \ref{opt:indirect} contain non-linear functions with integer variables. To solve these problems, we search for all possible configurations within the parameter ranges indicated in the above problem formulation.

\section{Simulation Results}
\label{sec:results}

In this section, we present and discuss the results of the optimization problems introduced in the previous section via link-level and system-level simulations. We carried out studies using a $28$ GHz band comprising $400$ MHz of operating bandwidth. For the link-level analysis, the path loss exponents for the \gls{bh} and \gls{ac} links are assumed to be $2$ and $3.2$. Our optimization considers a fully loaded \gls{gnb}, thus assuming full buffer traffic. The system-level study contains a multiuser scenario with stationary \glspl{ue}. We show results for two cases of fixed and varying PA efficiency separately.

\begin{figure*}[t]
    \centering
    \hspace*{\fill}%
    \begin{subfigure}[b]{0.3\textwidth}
        \centering
        \resizebox{\linewidth}{!}{\input{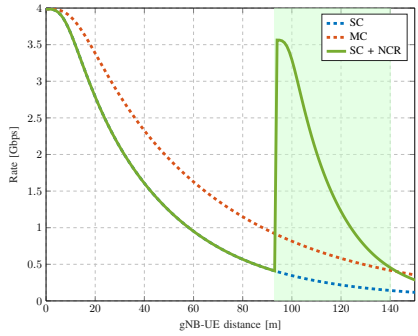}}
        \caption{Rate as a function of \gls{gnb}-\gls{ue} distance.}
        \label{fig:se_mc_sc}
    \end{subfigure}%
    \hspace*{\fill}%
    \begin{subfigure}[b]{0.3\textwidth}
        \centering
        \resizebox{\linewidth}{!}{\input{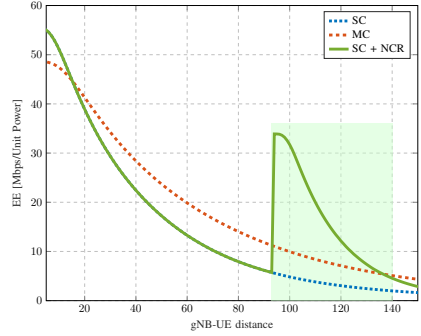}}
        \caption{\Gls{ee} as a function of \gls{gnb}-\gls{ue} distance.}
        \label{fig:ee_sc_mc}
    \end{subfigure}
    \hspace*{\fill}%
    \caption{Link-level simulation results for the comparison \gls{sc} vs. \gls{mc} vs. \gls{sc} + \gls{ncr} (assuming fixed \gls{pa} efficiency).}
    \label{fig:sc_mc_all}
\end{figure*}

\subsection{Link-level Direct Topology}
\label{subsec:linklevel_direct}

First, we present the \gls{ee} optimization results for the \emph{direct} topology of Fig.~\ref{fig:simple_top}. Practically, this case explores whether the most spectral efficient configuration of a \gls{gnb} always optimizes \gls{ee} and \gls{nes} for a \gls{ue} location. If not, we are seeking the best \gls{gnb} parameters adaptation strategy to achieve \gls{nes} and higher \gls{ee}.


Fig.~\ref{fig:direct_overall} shows the link-level optimization results for the \emph{direct} topology (\ref{opt:direct}). \Gls{ee} of the \gls{ee}-optimal configurations, relative to the baseline configuration (which assumes the maximum value for each \gls{gnb} parameter - highest \gls{se}), are shown for the fixed and varying \gls{pa} efficiency cases. Furthermore, the impact of \gls{ee} optimization on the \gls{ue} rate is provided to clarify the trade-off between \gls{se} and \gls{ee}.

There is an opportunity for \gls{nes} only in high \gls{snr} regimes (i.e., small \gls{gnb}-\gls{ue} distance, e.g. less than $35$ meters as highlighted in green in Fig.~\ref{fig:direct_overall}). At larger \gls{gnb}-\gls{ue} distances, the impact of \gls{snr} reduction due to a less spectral-efficient \gls{gnb} configuration is greater than the corresponding $P_{gNB}(DL)$ reduction, excluding any possibility of \gls{nes}. 

Regarding the \gls{ee}-optimal strategy, it can be shown that the highest available bandwidth is always \gls{ee}. However, when there is room to back off from the maximum configuration (that is, \gls{se} is compromised), the optimal strategy to achieve \gls{ee} operation would be different according to the \gls{pa} efficiency model. With fixed \gls{pa} efficiency, the optimal strategy is to first reduce the \gls{pa} output power $PAout_{gNB}$, and then (if there is still room for energy savings) adapt the number of active transmitting antenna elements $N_{Tx,gNB}$. The rationale behind this is that reducing the number of active transmitting antennas is more costly than reducing $PAout_{gNB}$ from a \gls{se} perspective ($SNR \propto PAout_{gNB} \cdot N_{Tx,gNB}^2$). In contrast, reducing $N_{Tx, gNB}$ or $PAout_{gNB}$ allows for the same level of \gls{nes} ($P_{gNB}(DL) \propto PAout_{gNB} \cdot N_{Tx,gNB}$). 
With varying \gls{pa} efficiency, the optimal \gls{ee} strategy is to first reduce the number of active transmitting antennas until it reaches a minimum value and then reduce $PAout_{gNB}$. This is because, in terms of \gls{nes}, reducing $PAout_{gNB}$ may not provide significant savings (since $P_{gNB}(DL) \propto \frac{PAout_{gNB}}{\eta(PAout_{gNB})}$ and, for varying \gls{pa} efficiency, this may be approximately constant). 

In conclusion, we see that for fixed PA efficiency, the network can achieve up to 30\% EE improvement with a maximum 10\% rate degradation. For varying PA efficiency, \gls{ee} gains can reach 60\%, but with a potential $\approx$20\% SE decrease.



\subsection{Link-level Indirect Topology}
\label{subsec:linklevel_indirect}

For the \emph{indirect} topology of Fig.~\ref{fig:repater_top}, we want to assess whether a higher network \gls{ee} can be achieved by backing off from the most \gls{se} configuration on either or both the \gls{bh} and \gls{ac} links. If this is the case, what parameters of \gls{gnb} and \gls{ncr} will mainly influence \gls{nes} and \gls{ee}?

Fig.~\ref{fig:indirect_overall} shows the results of the link-level optimization for the \emph{indirect} topology (\ref{opt:indirect}).
It can be seen that \gls{ee} optimization greatly increases energy efficiency compared to baseline (+15\%/56\%), with a slight degradation of rate performance (-3\%/20\%) only in high \gls{snr} regimes, as highlighted in green in Fig.~\ref{fig:indirect_overall}. To understand why, we note that the $effSNR(DL)$ of Eq.~\ref{eq:indirect_snr}, can be rewritten as follows:
\begin{align}
    effSNR(DL) &= \frac{SNR_{AC}}{1 + \frac{1 + SNR_{AC}}{SNR_{BH}}}.
    \label{eq:indirect_snr_alt}
\end{align}
A closer examination of the above formula reveals that if $SNR_{BH} >> 1 + SNR_{AC}$, then $effSNR(DL) \approx SNR_{AC}$. That is, if \gls{bh} \gls{snr} is much stronger than \gls{ac} \gls{snr} (more precisely, $1 + SNR_{AC}$), then the $effSNR(DL)$ will always be bottlenecked by $SNR_{AC}.$ This is in fact typically the case in practice, due to the more favorable \gls{bh} channel conditions compared to the \gls{ac} channel, as well as more antenna elements on an \gls{ncr} compared to a \gls{ue}. Now, when solving the optimization problem (Eq.~\ref{eq:ee_indirect}), one can typically reduce $SNR_{BH}$ without any significant impact on $effSNR(DL)$. This provides great opportunities for \gls{gnb} power savings (that is, reducing $P_{gNB}(DL)$), and therefore improving \gls{ee}. 

\begin{table}[t]
    \centering
    \begin{tabular}{|>{\raggedright}m{0.8cm}|>{\raggedright}m{1cm}|>{\raggedright}m{0.8cm}|>{\raggedright}m{0.7cm}|>{\raggedright}m{0.7cm}|>{\raggedright}m{0.8cm}|>{\raggedright}m{0.7cm}|}
    \hline
    \multicolumn{2}{|c|}{} & \multicolumn{2}{c|}{$gNB$} & \multicolumn{3}{c|}{$NCR$} \\
    \hline
    \multicolumn{2}{|c|}{Scheme} &\centering $PAout$ & \centering $N_{Tx}$  &\centering $N_{Rx}$ &\centering $PAout$ & \centering $N_{Tx}$
    \tabularnewline \hline
    \multicolumn{2}{|c|}{Baseline (Max conf)} &\centering  5.167 & \centering 192 & \centering 32 & \centering 13 & \centering 32
    \tabularnewline \hline
    \multirow{3}{0.8cm}{EE Optimal} & Low PL (15 m) & \centering{\cellcolor{yellow!30}} 0 & {\cellcolor{yellow!30}} \centering 76 & {\cellcolor{yellow!30}} \centering 20 & {\cellcolor{green!20}}\centering 12 & \centering 32        
    \tabularnewline \hhline{|~*{6}{-|}}
    & Med PL (24 m) &\centering {\cellcolor{yellow!30}} 0 & {\cellcolor{yellow!30}} \centering 60 & {\cellcolor{yellow!30}} \centering 18 & \centering 13 & \centering 32 
    \tabularnewline \hhline{|~*{6}{-|}}
    & High PL (48 m) &\centering {\cellcolor{yellow!30}} 0 & {\cellcolor{yellow!30}} \centering 48 & {\cellcolor{yellow!30}} \centering 12 & \centering 13 & \centering 32 
    \tabularnewline \hline
    \end{tabular}
    \caption{Optimal parameters at different path loss regimes for the \emph{indirect} topology with fixed \gls{pa} efficiency ($PAout$ is expressed in dBm). Between parenthesis is the \gls{ncr}-\gls{ue} distance in meters.}
    \label{tab:ind_results_fixed}
    \vspace{-4mm}
\end{table}

Table~\ref{tab:ind_results_fixed} shows the optimal choice of \gls{gnb} and \gls{ncr} parameters for a few path loss regimes, for the case of fixed \gls{pa} efficiency. As highlighted in yellow, the weaker the \gls{ac} link (higher path loss), the more we can save on the \gls{bh} side by reducing the \gls{gnb} Tx and \gls{ncr} Rx configurations. Consequently, more \gls{nes} can be achieved for increased \gls{ncr}-\gls{ue} distance.
The general strategy for \gls{ee} operation and \gls{nes} in the \emph{indirect} topology with fixed \gls{pa} efficiency is to first perform the adaptation of $PAout_{gNB}$ and, second, if more back-off is available, perform the adaptation of the number of \gls{gnb}'s active transmitting and \gls{ncr}'s active receiving antenna elements. Highlighted in green in Fig.~\ref{fig:ee_snr_indirect_fixed} and Table~\ref{tab:ind_results_fixed}, we see that for low \gls{ac} path loss values, it is more \gls{ee} to reduce the maximum \gls{eirp} on the \gls{ncr}'s \gls{ac} side via $PAout_{NCR}$ adaptation.

In the case of varying \gls{pa} efficiency, the optimal \gls{ee} strategy involves first adapting the active transmitting antennas of \gls{gnb} and the active receiving antennas of \gls{ncr}. If more back-off is allowed, further energy savings can be realized through reduction of $PAout_{gNB}$. Table~\ref{tab:opt_strategies} provides a summary of \gls{nes} strategies for each topology and \gls{pa} efficiency type.

\begin{table}[ht]
    \centering
    \begin{tabular}{|>{\raggedright}m{1cm}|>{\raggedright}m{3.2cm}|>{\raggedright}m{3.2cm}|}
    \hline
     & \multicolumn{2}{c|}{Topology} \\
    \hline
    \centering\Gls{pa} type & \centering Direct & \centering Indirect
    \tabularnewline \hline
    \centering Fixed Eff. & \centering Reduce $PAout_{gNB}$. If more room for back off, reduce $N_{Tx,gNB}$ & \centering Reduce $PAout_{gNB}$. If more room for back off, reduce $N_{Tx,gNB}$ and $N_{Rx,NCR}$
    \tabularnewline \hline
    \centering Varying Eff. & \centering Reduce $N_{Tx,gNB}$. If more room for back off, reduce $PAout_{gNB}$ & \centering Reduce $N_{Tx,gNB}$ and $N_{Rx,NCR}$. If more room for back off, reduce $PAout_{gNB}$
    \tabularnewline \hline
    \end{tabular}
    \caption{Summary of best \gls{ee}/\gls{nes} strategies for both \gls{pa} efficiency types and each topology. More room for back off comes at the cost of \gls{se}.}
    \label{tab:opt_strategies}
    \vspace{-4mm}
\end{table}

\subsection{Direct vs. Indirect Topology}

This subsection compares two topologies in terms of \gls{ee} and \gls{se}. We aim to determine whether it is more energy-efficient to achieve target coverage using a \gls{gnb} (e.g., a small cell) with an assisting \gls{ncr}, or a single yet stronger \gls{gnb} (e.g., a macro cell). We adopt the following methodology for this analysis.
A target \gls{dl} \gls{snr} (for example, $0$ dB) is set for \glspl{ue}. In a first topology (\gls{sc} + \gls{ncr}), we start by determining the direct coverage range of the \gls{sc} (i.e. a distance from the \gls{sc}, where a \gls{ue} would achieve the target \gls{snr} directly from the \gls{sc}), and place the \gls{ncr} at the edge of the \gls{sc} coverage region ($d_{BH}=94$ m). Next, the coverage range of the \gls{ncr} is calculated (that is, the distance from the \gls{ncr} where \glspl{ue} would achieve the $effSNR(DL)$ target \gls{snr} from the \gls{ncr} -- $d_{AC}=49$ m). In a second topology, we choose the smallest \gls{mc} (in terms of the number of transmitting antennas) that can provide the same coverage range (i.e., $d_{BH}+d_{AC}$) as that of the first topology. The two topologies are then compared in terms of \gls{ee} and \gls{se}.

Fig.~\ref{fig:se_mc_sc} shows that \gls{sc} + \gls{ncr} provides a much better rate in the \gls{ncr}'s coverage area (highlighted in green), that is, on average more than 2.6X the \gls{mc}'s \gls{se} in the same region. Outside of this region, \gls{sc} provides a less spectral-efficient service, 26\% reduced on average compared to \gls{mc}.

Fig.~\ref{fig:ee_sc_mc} shows the \gls{ee} comparison between the two topologies. As confirmed by the region highlighted in green, \gls{sc} + \gls{ncr} offers 2.1X better \gls{ee} on average in the \gls{ncr}'s coverage area. Outside of \gls{ncr} coverage, \gls{sc} provides slightly lower \gls{ee} compared to \gls{mc}, (-15\% on average). 

We can conclude that \glspl{ncr} are a promising solution (both in terms of spectral and energy efficiencies) to provide a target coverage. Especially if there is a hotspot region, a repeater will be able to improve the experience of \glspl{ue} in that region at a lower network energy consumption.  

\subsection{System-level}

\begin{figure}[t]
    \centering
    \includegraphics[width=0.62\linewidth]{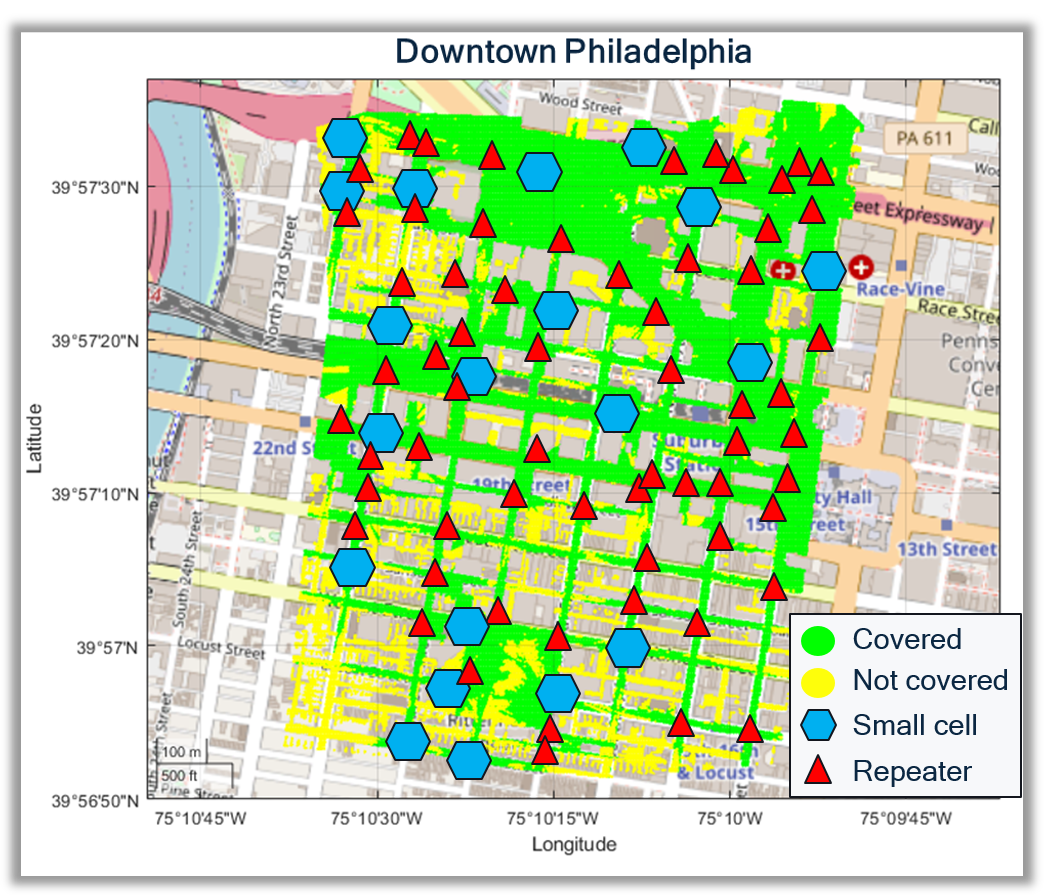}
    \caption{28 GHz Network based on a Downtown Philadelphia neighborhood \cite{Ahmed} (Map data © OpenStreetMap contributors, Microsoft, Esri Community Maps. License: https://creativecommons.org/licenses/by-sa/2.0/legalcode).}
    \label{fig:philly}
\vspace{-4mm}
\end{figure}

In the system-level analysis, we consider a realistic deployment of 20 \glspl{sc} and 62 \glspl{ncr} in the downtown area of Philadelphia, as shown in Fig.~\ref{fig:philly}. This deployment is the result of an optimization process that aims to find the optimal number and location of \glspl{sc} and \glspl{ncr} to achieve a target coverage of $80\%$ in \gls{fr}-2 while minimizing the deployment cost given a specific geographical region. Each \gls{sc} comprises three sectors and each sector might have one or more \glspl{ncr} to extend network coverage. Details for this deployment optimization are provided in \cite{Ahmed}.
From the optimal deployment, we consider two topologies: one comprising both \glspl{sc} and \glspl{ncr} (\emph{With Repeaters} in the figures) and the other comprising only \glspl{sc} (addressed as \emph{Without Repeaters}). For comparison between the two topologies, we focus on the reduced set of \gls{ue} locations that are covered in the topology \emph{Without Repeaters}. This is because certain \glspl{ue}, which are served via an \gls{ncr} from the optimization solution, can also be served directly with a \gls{sc} at lower \gls{snr}. With this system-level study, we would like to confirm the findings of the link-level analysis.
Furthermore, we want to measure the impact \glspl{ncr} have in terms of \gls{ue} throughput and \gls{ee}/\gls{nes} performance at the system level.

\begin{figure*}[t]
    \centering
    \hspace*{\fill}%
    \begin{subfigure}[b]{0.31\textwidth}
        \centering
        \resizebox{\linewidth}{!}{\input{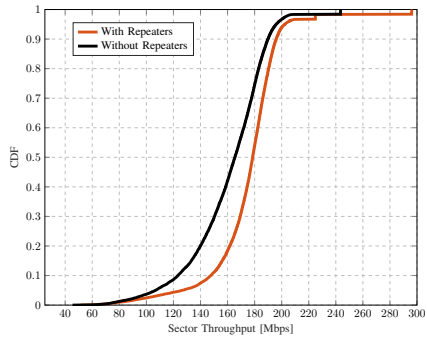}}
        \caption{CDF of the \gls{dl} throughput per sector for the topology solution \emph{With Repeaters} against the topology \emph{Without Repeaters}.}
        \label{fig:sys_tput_sector}
    \end{subfigure}%
    \hspace*{\fill}
    \begin{subfigure}[b]{0.31\textwidth}
        \centering
        \resizebox{\linewidth}{!}{\input{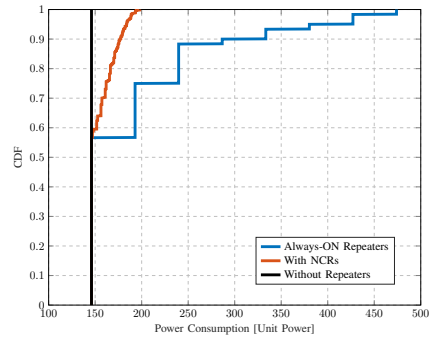}}
        \caption{CDF of the power consumption per sector. The \gls{ncr} can be smartly turned on/off based on whether it is serving indirect \glspl{ue}.} 
        \label{fig:sys_power_varying}
    \end{subfigure}
    \hspace*{\fill}%
    \vspace{4mm}
    \hspace*{\fill}%
    \begin{subfigure}[b]{0.31\textwidth}
        \centering
        \resizebox{\linewidth}{!}{\input{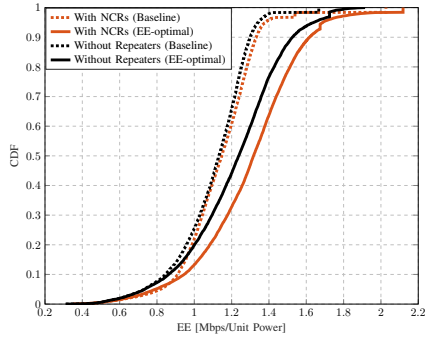}}
        \caption{CDF of the \gls{ee} per sector. In the baseline results, \glspl{sc} and \glspl{ncr} use the most \gls{se} configuration, while the EE-optimal results come after optimizations (\ref{opt:direct})-(\ref{opt:indirect}).}
        \label{fig:sys_ee_sec_varying}
    \end{subfigure}%
    \hspace*{\fill}
    \begin{subfigure}[b]{0.31\textwidth}
        \centering
        \resizebox{\linewidth}{!}{\input{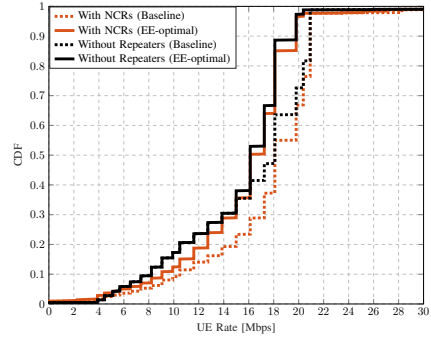}}
        \caption{CDF of the \gls{ue} \gls{dl} rate. In the baseline results, \glspl{sc} and \glspl{ncr} use the most \gls{se} configuration, while the EE-optimal results come after optimizations (\ref{opt:direct})-(\ref{opt:indirect}).}
        \label{fig:sys_ue_rate}
    \end{subfigure}
    \hspace*{\fill}%
    \caption{System-level simulation results based on the deployment shown in Fig.~\ref{fig:philly}. Each \gls{sc} has 3 sectors, therefore, 60 sectors are present in the deployment.}
    \label{fig:sys_overall}
\end{figure*}

Fig.~\ref{fig:sys_tput_sector} shows that \glspl{ncr} can help \glspl{sc} deliver a higher overall throughput, because they enable stronger connections to \glspl{ue} and serve them with higher \gls{se}. However, this comes at the cost of increased network power consumption, as can be seen in Fig.~\ref{fig:sys_power_varying}. Note that almost 60\% of the sectors do not have any associated \glspl{ncr} and therefore consume the same power in the two different topologies considered. For sectors with associated repeaters (\emph{With Repeaters} topology), having always-on repeaters increases the power consumption by around 65\%. However, deploying intelligent \glspl{ncr} (that can be turned on and off by the network - for example according to the current \gls{ue} schedule) shows to significantly reduce the overall network power consumption, although still around 15\% higher than a topology \emph{Without Repeaters}. 

From link-level studies, we saw that \gls{ee}-optimal strategies can be adopted to improve the energy efficiency of the network. At the system level, for the topology \emph{With Repeaters}, we observe that there is a great opportunity for \glspl{nes} through the \gls{bh} links. Fig.~\ref{fig:sys_ee_sec_varying} shows the CDF of the \gls{ee} per sector with \gls{ee} optimization and without optimization (i.e., the baseline with highest \gls{se}) for the case of varying \gls{pa} efficiency. The \gls{ee}-optimal topology \emph{With Repeaters} outperforms the \gls{ee}-optimal topology \emph{Without Repeaters} in terms of the overall \gls{ee}, with an average increase of 8.5\%, and the baseline topology \emph{With Repeaters}, with an average increase of 18\% in \gls{ee}. The impact of \gls{ee} optimization on the performance of \glspl{ue} can be seen in Fig.~\ref{fig:sys_ue_rate}. Specifically, the figure shows that a topology \emph{With Repeaters} (after \gls{ee} optimization) can provide better coverage for the cell edge \glspl{ue}, with an average 5\% \gls{dl} rate improvement compared to the topology \emph{Without Repeaters}. At the same time, the \gls{ue} rate suffers a slight 9\% degradation on average compared to the baseline topology \emph{With Repeaters}. Thus, \gls{ee} optimization brings considerably higher \gls{ee} and \gls{nes} at the expense of slightly degraded \gls{ue} performance, as we found in our analytical study. 

\section{Conclusions}
\label{sec:conclusion}

In this paper, we introduced the challenge of optimizing the network \gls{ee} in the context of future \gls{6g} networks enabled by \glspl{ncr}. We presented \gls{nes} strategies to improve \gls{ee}, thus decreasing network power consumption, by adapting the transmitting and/or receiving configurations of \gls{gnb} and \gls{ncr}. Our link and system-level results show that it is more energy-efficient to sacrifice \gls{se} in high \gls{snr} regimes. To this end, the best strategy depends on the \gls{pa} implementation based on the availability of bias adjustment. Bias adjustment represents a more advanced \gls{pa} technology that improves \gls{ee} compared to previous/legacy \gls{pa} technologies. We showed that repeaters guarantee a higher overall \gls{se} compared to a deployment without repeaters and allow power savings at \gls{gnb} through their typically strong \gls{bh} link, leading to a higher overall \gls{ee}. Therefore, repeaters represent an energy- and cost-efficient solution to increase the coverage and capacity of future \gls{6g} networks.
Finally, despite our work focuses on \gls{fr}-2, the optimization results can be exploited for \gls{fr}-3, which is a candidate spectrum targeted for future \gls{6g} deployments.

\bibliographystyle{IEEEtran}
\bibliography{bibl}


\end{document}